# A Naive Approach for Automatic Line-level Code Completion

Shamima Naznin
Department of Computer Science and Engineering,
Bangladesh University of Engineering and Technology
Dhaka, Bangladesh
shamimanaznin21@gmail.com

Manishankar Mondal
Computer Science and Engineering Discipline,
Khulna University
Khulna, Bangladesh
mshankar@cseku.ac.bd

*Abstract*—Coding is an integral aspect of programming. A programmer can automatically complete a code fragment after writing a few tokens and the process of automatically completing a code fragment is known as code completion. A number of research on code completion have previously been conducted for method body completion and method parameter completion. However, this fundamental study explores the idea of automatically completing any program statement that might not even be part of a method. The goal of this study is to provide a number of suggestions to the programmer for completing code throughout the codebase by identifying and analyzing code similarities. The proposed methodology of this study can be regarded as a fundamental framework for code completion in the field of automated code completion. From the investigation of hundreds of revisions of four subject systems written in C and Java, it is observed that the proposed method can automatically complete around 22% of code statements with an average accuracy of 87% that a programmer writes during development which will accelerate software development time. This paper's empirical analysis further demonstrated that the proposed approach can be used with programming language neutrality. The study concludes by illustrating that taking 10 characters as prefixes before invoking completion provides maximum precision.

## I. INTRODUCTION

Code completion is a promising research area that has been explored much in the last few decades. According to the literature, code completion refers to the mechanism of automatically completing an incomplete code fragment written by a programmer. When a programmer writes the first few tokens of a code fragment, a code completion mechanism provides the programmer with a number of suggestions for completing the code. Code completion is the most appealing feature of a modern integrated development environment (IDE).

Code completion is the most frequently used tool in software development [1]. Murphy [2] et.al showed that after a basic operation like a copy-paste and delete, code completion is the most used feature in Eclipse IDE. A number of techniques have been introduced to support code completion in modern IDEs. For instance, Visual Studio, a popular C# code editor environment, supports variable completion, method call completion of an object, method body completion, class template instantiations, pattern completion based on a context etc [3]. In search of an easy-going yet effective approach to code completion, a technique is introduced here that provides suggestions and completion of a whole line of code based on textual similarity from the current code base. We proposed a code completion approach that utilizes recent code changes to provide more relevant suggestions. This is particularly beneficial for developers working on complex and dynamic codebases. The proposed method uses simple string matching to find appropriate code completion suggestions to complete any line of code. Additionally, a rigorous study is carried out on the effectiveness of considering context while providing suggestions. This empirical study involves the three research questions outlined in Table 1.

TABLE I: Research Questions

| SL | Research Question |
|---|---|
| RQ 1 | Is it possible to provide line-level code suggestions/completion using the current code base of a software system? |
| RQ 2 | What is the precision for providing suggestions to complete any statement? |
| RQ 3 | How many number of characters can be considered at the time of providing suggestion for completing any line of code? |

## II. RELATED WORKS

Different code completion techniques provide support for different types of program statements [4], [5], [6], [7], [8], [9], [10], [11]. However, the capability to provide code completion for all types of statements would be more convenient which is the motivation behind this study. The most widely used Eclipse IDE supports template-based API completion [12]. Eclipse's code completion engine supports various completion techniques, including method call completion, constructor completion, override completion, subtype completion, sub-word completion, and adaptive template completion [12]. Nguyen et al. [13, 14] described a technique called SLAMC (Statistical Semantic Language Model for Code) that provides code completion and suggestions. It operates on the semantic level and is based on a semantic token. In another study, Kim et al. [15] introduced a tree-based and Nguyen introduced a [16] graph-based approach (GraPacc) that extracts context-sensitive features from the code under editing. Hill and Rideout [17, 18] described an idea of code completion of the body of a method by employing a machine learning algorithm for finding near duplications/clones at the fragment level. Bruch et al. [19] presented a system for intelligent code completion that acquires knowledge from existing code repositories. On the basis of examples in the database, he demonstrated three code completion techniques: frequency-based code completion (FCC), association rule-based code completion (ArCC), and best matching neighbor code completion (BMN) to suggest the method call for a single variable under editing. CSCC (Context Sensitive Code Completion) [20] focused on automatic method call completion. It used a frequency of method calls for rank-ing and suggestion based on context-sensitive features. Robbes and Lanza [21] used entire changes in program history for code completion. Hindle et al. [22] showed that n-gram probabilistic model-based code completion is much more flexible and it can suggest any next token using type information from the current IDE. An n-gram probabilistic model uses a frequency of tokens for ranking and providing a list of suggestions. C Zhang [23] proposed an approach named "Precise" to im-prove developers' productivity by performing code completion through the automatic prediction of parameters for any method called completion. It uses the KNN algorithm for finding the instance that is most similar to the current context. and uses the type and structure information of that instance for the recommendation of the actual parameter. "Cookbook" [24]-a new code completion technique that allows developers to define a custom, reusable template for complex edit operations by clearly identifying change examples. Cookbook performs an AST-based operation for matching and editing operations and saves them for later use.

## III. BACKGROUND AND PROPOSED METHOD

### TABLE II: Subject Systems

| Systems | Language | Domains | LLR | Revisions |
|---|---|---|---|---|
| Ctags | C | Index Generator | 33,270 | 774 |
| BRL-Cad | C | 3D Modeling | 39,309 | 735 |
| Jabref | Java | Reference Management | 45,515 | 1545 |
| Java-ML | Java | Machine Learning Library | 16,428 | 1200 |

The proposed method of this empirical study is intended to work for any software system, hence, diverse subject systems from different domains from two programming languages are chosen so that the approach remains valid in case of system changeability. For investigating the proposed approach, the software systems of Table (II) downloaded from an online SVN repository[25] that are written in C and Java. Here, a code completion method is designed in such a way that could search the current codebase and provide suggestions. This completion method completes a whole line of code by detecting code similarity. For finding code similarity here used a naive approach which is a string-matching technique. As suggestions are provided by detecting code similarity, it is expected that the proposed approach to be lightweight and faster in comparison with the existing mechanisms.

The schematic diagram presented above shows an overview of the code completion process we introduced here.

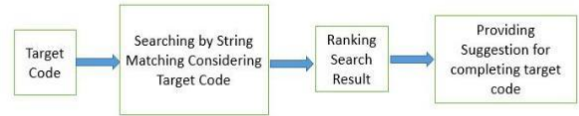

Fig. 1: Schematic Diagram for Our Proposed Method.

### A. Target Code

In our code completion method, the target code is the code that the programmer is currently writing. This is the code that the programmer wants to complete, and it is represented by the prefix code that has already been written.

### B. Search code Similarity

Once the target code has been identified, the proposed method searches the current codebase for lines of code that begin with the same characters. Since this search may return multiple results, a ranking mechanism is used to rank the results. The recency ranking mechanism applied here, which was successfully used by Omori et al. [26], to rank our search results.

### C. Finding Line Addition

To use this code completion approach, we need to handle the additional lines that provide suggestions for code completion. We utilized the diff tool [26] to compare the source files of two consecutive versions. If any lines are added, we will locate those lines in a database (Added Line Database) along with their revision identity.

### D. Checking Added Line

Using string matching (String matching is a technique for finding one or more occurrences of a substring within a string and is considered the fastest and lightweight matching mech-anism), here checked whether the previous revision contained the added lines of the Added Line Database, i.e., whether suggestions can be provided from the previous version.

### E. Ranking

Then the search results are stored in a new database and ranked according to the most recent occurrence among all previous versions.

### F. Providing suggestions for target code

In order to provide a suggestion for the target code, it is needed to calculate precision by indicating the number of true and false suggestions among the total number of suggestions that are recommended and located in the suggestion database.

### G. Proposed Methodology

Our proposed methodology serves as a fundamental framework that is applicable to a wider spectrum of software systems, and it can be considered as a groundwork for exploring machine learning or deep learning techniques.

The methodology we intended to develop was formulated through an investigation of software systems outlined in Table II, all of which were sourced from an online SVN repository [25].

Our exploratory study encompasses the following key steps:

1) Data Collection: In the initial step of our investigation, we downloaded all revisions of the software systems specified in Table II directly from the online SVN repository [25]. An SVN repository is a central storage location for files and directories with their revision history. [27].
2) Source Code Comparison: Subsequently, the UNIX diff tool [28] was employed to compare the source files between two consecutive versions (Figure 2). The primary objective during this phase is the identification of lines that were newly added in the consecutive version's source code, with a particular focus on added lines. Lines that were deleted or modified were excluded from our analysis.
3) Added Line Database: To facilitate our analysis an "Added Line Database" was established that functions as a comprehensive repository for these added lines (Figure 2). This database includes both the lines themselves and their corresponding version numbers.
4) String-Matching Technique: Our study then focused on investigating the presence of added lines from the (i+1)-th revision in the i-th revision. We accomplished this by employing an SVN repository string-matching technique, enabling us to make detailed comparisons with the source code from the previous revision. This step served as the baseline for assessing the feasibility of providing code completion suggestions for the added lines in the i-th revision.
5) Precision Evaluation: Finally, precision was evaluated by scrutinizing the accuracy and effectiveness of the code completion suggestions generated for each added code line.
   This empirical investigation aimed to provide valuable insights into the recall and precision of our line-level code completion approach.

## IV. EVALUATION METRIC

In order to evaluate the effectiveness of our method, we consider the following three evaluation metrics.

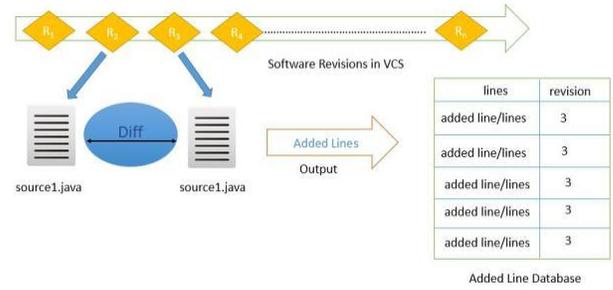

Fig. 2: An overview of source code comparison and added line from the proposed methodology.

Recall: The recall is the percentage of the number of actual suggestions retrieved with respect to the total number of actual suggestions[29]. A mechanism with a recall value of 100% indicates that it can provide all actual suggestions. The following equation can express recall.

$$Recall = \frac{|Retrieved\ Relevant\ Suggestions|}{All\ Relevant\ Suggestions} \times 100 \quad (1)$$

Precision: Precision is the percentage of the number of actual suggestions provided with respect to the total number of suggestions provided [29]. Having 100% precision means that all the results retrieved by a mechanism are relevant. The following equation can compute it.

$$Precision = \frac{|Retrieved\ Relevant\ Suggestions|}{|All\ Retrieved\ Suggestions|} \times 100 \quad (2)$$

F1 Measure: Precision and recall scores are not discussed in isolation. Instead, both are combined into a single measure. This measurement that combines precision and recall by calculating the harmonic mean of precision and recall is known as F-measure or F1 measure or F1 score [29]. The equation for the F1 measure is given below.

$$F1measure = \frac{2 \times precision \times recall}{precision + recall} \quad (3)$$

## V. EXPERIMENTAL RESULTS ANALYSIS

Answering research questions is the paramount goal of our research. In this section, we attempt to answer the research questions by analyzing the experimental results.

### A. Answering RQ 1

Is it possible to provide line-level code suggestion/completion using the current code-base of a software system?

To answer this, we performed step (3) in our experimental steps. From our experimental results (table III), we found that for JavaML, out of 1046 added lines, we can suggest 327 lines of code from the current working code-base. For Ctags the number of suggestible lines is 158 out of 918, for BrlCad it is 341 out of 1555, and for Jabref it is 606 out of 3378. For the

JavaML system, we can provide line completion suggestions up to 31% (Recall) using the current code-base. Other subject systems of our experimental procedure also demonstrate a noteworthy recall. The average recall for the four subject systems is around 22%. This signifies that we can perform line-level code completion for a software system using the current codebase on an average of 22%. The details of our findings about the subject system are listed in Table(III).

### B. Answering RQ 2

What is the precision for providing suggestions to complete any statement?

Precision refers to the availability of relevant suggestions in the suggestion list for a target code. To find the overall precision of the four subject systems, we conducted the following steps. First, we fetched each added line from the A_Database (added line database). We call this single added line a code completion candidate. Second, for each added line we extracted a certain number of characters from that line. We use this line as the target code in which we invoke our code completion. The remaining part of the added line is used to verify the suggestions given by the code completion mechanism. The number of characters that provide good precision and recall has been investigated further in the following section. After taking the target code line, we searched for the code line in the code base that has the exact same starting characters as the target code. We extracted these code lines and provided suggestions for completing the target code. Third, we calculated how many of those suggestions match with the code completion candidate (relevant suggestion) and how many of those don't match (irrelevant suggestion). Then we calculated precision with equation (2). The findings of our experiment with subject systems are shown in the table (IV).

### C. Answering RQ 3

How many characters can be considered at the time of providing suggestions for completing any line of code?

A comparative precision and recall chart with F measure is shown in fig (3). In this figure, the vertical line represents the percentage of successful code completion and the horizontal line denotes the number of characters in the target code. Our empirical study with four subject systems indicates that the number of characters chosen as target code affects precision and recall. Before invoking code completion, a particular number of characters must be provided. The table shows how varying character counts affect precision and recall. Increased target code characters increase precision and decrease recall. Adding characters to target code reduces irrelevant codebase suggestions. For the same number of target code characters, C-based systems are more precise than Java-based systems. Java has longer variables and methods than other structured languages. Its class structure requires large variable names. Therefore, we can say that software systems that use long naming conventions may have to provide additional characters prior to invoking code completion. A comparative precision and recall chart with F measure is shown in fig (3). In this figure, the vertical line represents the percentage of successful code completion and the horizontal line denotes the number of characters in the target code.

TABLE III: Recall for Subject Systems

| Systems | Total added line | Suggestable Line | Recall% |
|---|---|---|---|
| Ctags | 918 | 158 | 17.21 |
| BRL-Cad | 1555 | 341 | 21.92 |
| Jabref | 3378 | 606 | 17.94 |
| Java-ML | 1046 | 327 | 31.26 |

TABLE IV: Precision for Subject Systems

| Systems | Total added line | Suggestable Line | Precision(%) |
|---|---|---|---|
| Ctags | 918 | 158 | 90.02 |
| BRL-Cad | 1555 | 341 | 87.52 |
| Jabref | 3378 | 606 | 78.23 |
| Java-ML | 1046 | 327 | 78.36 |

TABLE V: Average precision and recall and F1-score for different characters

| Character | Recall(%) | Precision(%) | F1-Score(%) |
|---|---|---|---|
| 1 | 99.9725 | 0.9975 | 1.9575 |
| 2 | 98.8075 | 2.5925 | 5.03 |
| 3 | 94.23 | 5.3175 | 10.0125 |
| 4 | 90.335 | 9.45 | 17.0975 |
| 5 | 88.6175 | 14.575 | 24.8575 |
| 6 | 86.085 | 34.935 | 49.2425 |
| 7 | 82.19 | 55.0075 | 65.67 |
| 8 | 79.6875 | 68.7025 | 73.735 |
| 9 | 75.88 | 75.985 | 75.7525 |
| 10 | 71.71 | 83.5325 | 76.81 |
| 11 | 65.225 | 87.07 | 73.165 |

### VI. COMPARISON

A number of different state-of-the-art code completion methods have been investigated with various scenarios in order to evaluate the contribution of this study, table VI. We chose three different authors of code completion methods (CSCC[20], and two others[30],[15]) to compare with this paper's method. Though those techniques can perform various types of code completion with high precision, there are some scenarios where the method of this paper outperforms. In terms of language independence, Asaduzzaman[20], conducted empirical studies on a number of systems of multiple lan-guages of c and Java, which is similar to ours, whereas others worked with single language subject systems.

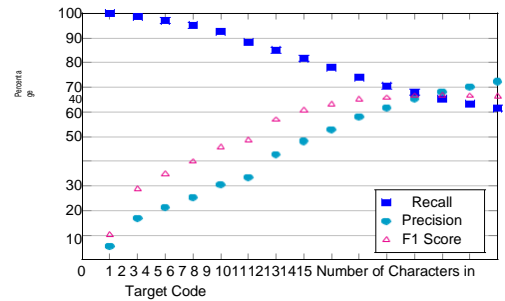

Fig. 3: Precision and Recall % for Different Number of Characters

Single variable consideration refers to providing suggestions for a single variable while typing. This is an important aspect of code completion for increasing programmer productivity by reducing the amount of manual typing and searching required to write code. The proposed method considers single variables for providing suggestions, while none of the other methods consider single variables.

At the time of code completion, when the parent node contains a single or multiple method calls as its parameter, CSCC[20], and other code completion mechanisms can't com-plete the whole code line in a single code completion invoca-tion. Rather it takes multiple code completion invocations to complete the code line. This multiple-time code completion invocation can result in requiring a longer time to develop a software system. On the other hand, our code completion mechanism can perform whole code line completion in a single invocation even when the code line contains a method that has method as its parameter.

In comparison with others, our proposed technique employs a lightweight comparison and ranking mechanism to make it suitable for large codebases. To verify its efficacy, we conducted a comprehensive empirical study on four subject systems that have undergone thousands of revisions. The results of the study demonstrate the ability of the technique to handle large codebases effectively. Additionally, our research highlights the importance of incorporating a prefix code of 10 characters when invoking code completion. Our findings indicate that the use of a prefix code before code completion can significantly enhance the F1 score while maintaining a high level of precision. This denotes the importance of considering both accuracy and efficiency when developing code completion techniques.

Hence, the method of this paper demonstrates its generality with its language independence, completion of a full line of code with a single invocation, ability to perform on a single variable, suitability for a large codebase, and considered cost sensitivity which uses a simple string matching technique rather than a more complicated one. We believe our method of code completion is not a replacement for the current code completion techniques rather it will effectively complement them.

## VII. CONCLUSION AND FUTURE WORKS

Providing line-level code completion and completion sug-gestions can accelerate the programmer's productivity and software system development. Our research has found that a typical software system faces a significant amount of recurring code lines. Up to 31% of code, lines can be completed with a single code completion invoked by matching the prefix code. Our method is both language-independent and lightweight in comparison to the current standard systems. There are already cutting-edge code completion mechanisms that address various types of code completion facilities. By completing the entire code line, our method of code completion is anticipated to be a helpful tool for those who require assistance with code completion. We believe that our method of code completion is not a replacement for existing code completion techniques, instead, it will augment them efficaciously. Future investiga-tion may entail applying machine learning, deep learning, and various techniques to improve the search results. Nevertheless, we will include a benchmark of the proposed method against existing code completion systems to demonstrate its perfor-mance improvements and advantages.

## VIII. THREADS OF VALIDITY

There is no database in the software systems that provides information regarding line addition, thus we cannot check it up. As a result, the number of line additions taken from the repositories may differ somewhat from the number of actual additions performed.

TABLE VI: Comparison with different methods

| Authors | | M Asaduz-zaman | Kim et. al | Wenhan Wang | Proposed method |
|---|---|---|---|---|---|
| Area of compari-son | Language Independency | Yes | No | No | Yes |
| | Single variable consideration | No | No | No | Yes |
| | Method hav-ing parameter | Yes | No | No | Yes |
| | Multiple invocation | Yes | Yes | No | No |
| | Character Consideration | No | Yes | Yes | Yes |
| | Large code base consideration | Yes | No | No | Yes |
| | Cost Sensitive | No | No | No | Yes |